\documentclass[journal=ascecg,manuscript=article,articletitle=true]{achemso}

\usepackage[version=3]{mhchem} 
\usepackage[sort&compress,numbers,super]{natbib}
\usepackage{amsmath}
\usepackage{graphicx}
\usepackage{wrapfig}
\usepackage{longtable}
\usepackage[colorinlistoftodos]{todonotes}
\usepackage[colorlinks=true, allcolors=blue]{hyperref}
\usepackage{subcaption}
\usepackage{makecell}
\usepackage{multicol}

\newcommand{\todoajm}[1]{\todo[color=green]{\@AJM: #1}}
\newcommand{\todobmc}[1]{\todo[color=red]{\@BMC: #1}}
\newcommand{\todomch}[1]{\todo[color=blue]{\@MCH: #1}}

\author{Benjamin M. Comer}
\affiliation[Georgia Institute of Technology]{School of Chemical \& Biomolecular Engineering, Georgia Institute of Technology, 311 Ferst Dr NW, Atlanta, GA, 30318}
\author{Andrew J. Medford}
\affiliation[Georgia Institute of Technology]{School of Chemical \& Biomolecular Engineering, Georgia Institute of Technology, 311 Ferst Dr NW, Atlanta, GA, 30318}
\email{andrew.medford@chbe.gatech.edu}
\phone{+1 (404) 385-5531}

\title{Analysis of Photocatalytic Nitrogen Fixation on Rutile TiO$_2$(110)}

\abbreviations{DFT}

\begin{document}







\begin{abstract}
Photocatalytic nitrogen fixation provides a promising route to produce reactive nitrogen compounds at benign conditions. Titania has been reported as an active photocatalyst for reduction of dinitrogen to ammonia; however there is little fundamental understanding of how this process occurs. In this work the rutile (110) model surface is hypothesized to be the active site, and a computational model based on the Bayesian error estimation functional (BEEF-vdW) and computational hydrogen electrode is applied in order to analyze the expected dinitrogen coverage at the surface as well as the overpotentials for electrochemical reduction and oxidation. This is the first application of computational techniques to photocatalytic nitrogen fixation, and the results indicate that the thermodynamic limiting potential for nitrogen reduction on rutile (110) is considerably higher than the conduction band edge of rutile TiO$_2$, even at oxygen vacancies and iron substitutions. This work provides strong evidence against the most commonly reported experimental hypotheses, and indicates that rutile (110) is unlikely to be the relevant surface for nitrogen reduction. However, the limiting potential for nitrogen oxidation on rutile (110) is significantly lower, indicating that oxidative pathways may be relevant on rutile (110). These findings suggest that photocatalytic dinitrogen fixation may occur via a complex balance of oxidative and reductive processes.
\end{abstract}
\section{Introduction}

\begin{figure}[h]
\centering
\includegraphics[width=0.4\textwidth]{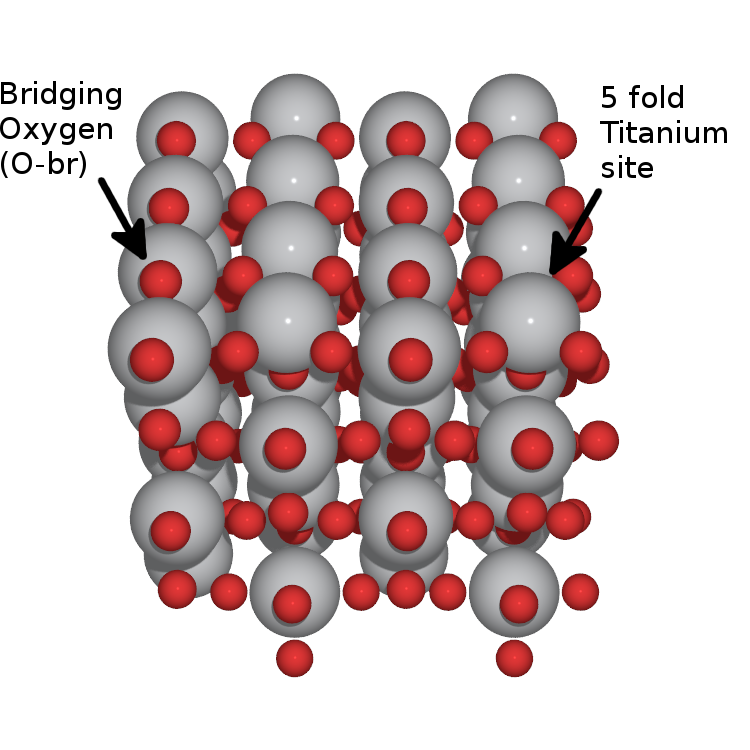}
\caption{Illustration of TiO$_2$ (110)}
\label{fig:model_surface}
\end{figure}

The photocatalytic reduction of atmospheric nitrogen was first proposed in the soil science community\cite{Dhar_1941}, and was later demonstrated under photocatalytic conditions by Schrauzer and Guth \cite{Schrauzer_1977,Schrauzer_1983}. These efforts spurred numerous studies on photocatalytic nitrogen fixation by titania catalysts \cite{Bickley_1979,Augugliaro_1982,Soria_1991,Schrauzer_2011, Yuan_2013,Hirakawa_2017, Medford_2017} that resulted in a variety of interesting observations. It was shown that dinitrogen can be converted to ammonia under both aqueous conditions \cite{Augugliaro_1982,Hirakawa_2017} and humidified air \cite{Schrauzer_1977,Schrauzer_1983}, with higher rates observed in gas-phase conditions \cite{Schrauzer_2011}. Further, the titania photocatalysts selectively reduce nitrogen to ammonia in most studies \cite{Schrauzer_1977,Schrauzer_1983,Augugliaro_1982,Soria_1991,Schrauzer_2011,Hirakawa_2017}, although some work indicates that oxidation to nitrates occurs \cite{Bickley_1979,Yuan_2013}, and other authors observed no photocatalytic nitrogen fixation activity \cite{edwards1992opinion,davies1993reply,Boucher_1995,Davies1995}. In addition, the presence of the rutile crystal phase and iron dopants were found to be critical for enhancing nitrogen reduction activity \cite{Schrauzer_1977,Schrauzer_1983,Augugliaro_1982,Soria_1991}. The iron dopants have been proposed to promote the formation of rutile domains \cite{Schrauzer_1977}, and/or enhance separation of charge carriers by acting as an electron sink \cite{Soria_1991}; while the exact role is not known, the prevailing hypothesis is that Fe is not part of the active site based on the observation that excess Fe content reduces catalytic activity \cite{Soria_1991}. More recently, the solar-to-ammonia efficiency was measured to be around 0.02\% for a pure titania sample, and the authors hypothesize that the reaction occurs via direct N-N bond scission at a rutile (110) bridging oxygen vacancy (O-br) active site.\cite{Hirakawa_2017} Despite the long history of photocatalytic nitrogen fixation on titania catalysts and numerous experimental results, there is little fundamental understanding of the molecular-scale processes that underly this important and potentially transformative photocatalytic process. In this work we investigate the hypothesis that rutile (110) (Figure \ref{fig:model_surface}) is the active surface for photocatalytic nitrogen fixation by examining the pristine rutile (110) surface, the defected (110) surface containing an oxygen vacancy, an iron substitution at the (110) surface, and oxidative processes on the pristine (110) surface.

The rutile TiO$_2$ (110) surface has been well-studied both experimentally \cite{Benkoula2015,Walle2009,Lu1994,Rusu2000,Rusu2001,Henderson_2011} and computationally \cite{Stodt2013,Sorescu2000,Cheng2011,Diebold_2003}; however, there have been relatively few computational studies of the interactions of rutile TiO$_2$ surfaces with nitrogen-containing compounds \cite{Stodt2013,Sorescu2000,Cheng2011,H_skuldsson_2017,Xie_2017}. Two comprehensive computational studies on nitrogen oxides were performed by Stodt et al. \cite{Stodt2013} and Sorescu et al. \cite{Sorescu2000}.  These studies examined N$_x$O$_y$ compounds on TiO$_2$ (110) surfaces both experimentally under ultrahigh vacuum (UHV) conditions and theoretically using density functional theory (DFT) at the generalized gradient approximation (GGA)\cite{Sorescu2000} and hybrid \cite{Stodt2013} levels of theory. This work showed that DFT is able to accurately obtain the vibrational frequencies of intermediates on the surface and is consistent with UHV experiments.
In addition, Cheng et al. \cite{Cheng2011} examined the adsorption of NH$_3$ (ammonia), NH$_2$, and H on clean rutile (110) surfaces using DFT. They concluded that NH$_3$ and NH$_2$ adsorb at the 5-fold titanium site, whereas hydrogen could bind at the bridging oxygens or the in-plane oxygen atoms. Several studies have also investigated the binding of ammonia and other nitrogen-containing species on anatase \cite{Onal_2006,Erdogan_2010,Erdogan_2011,Markovits_1996,Ji_2014} and monoclinic \cite{Guo_2012} polymorphs of TiO$_2$. Very recently Xie et. al. used DFT to calculate the energetics of nitric oxide reduction on TiO$_2$ \cite{Xie_2017}, and H\"{o}skuldsson et. al. utilized GGA DFT to screen rutile oxides for electrochemical ammonia synthesis, and found a relatively large limiting potential of ca. 2 V for the (110) surface of rutile TiO$_2$ \cite{H_skuldsson_2017}.

In contrast to the relatively low number of computational studies there have been many experimental studies of nitrogen compounds on rutile surfaces. Yates and colleagues conducted several studies of NO$_{\mathrm{x}}$ molecules under UHV conditions over rutile TiO$_2$ (110) \cite{Lu1994,Rusu2000,Rusu2001}. Their early work showed that NO adsorbs and decomposes to form N$_2$O on reduced surfaces \cite{Lu1994}, and a follow up study involved photochemical activation of NO on the surface that resulted in the production of N$_2$O, even with photon energies below the band gap \cite{Rusu2000}. When the surface is dosed with N$_2$O and exposed to photons at high coverage, N$_2$O is desorbed from the surface, while at low coverages N$_2$ is observed \cite{Rusu2001}. Much of this was also seen in the recent examination by Kim and colleagues with the additional insight that surface oxygen vacancies play an important role \cite{Kim2016,Kim2014}. Furthermore, numerous applied studies focusing on nitrogen doped titania \cite{Chen_2007}, TiO$_2$ as an ammonia sensor \cite{Karunagaran_2007,Suganuma_2015} or in the selective catalytic reduction (SCR) process \cite{Busca_1998,Ramis_1990,Giraud_2014,Liu_2017} provide insight into the interaction of titania with nitrogen-containing compounds. Although TiO$_2$ is typically not considered to be the active phase for the SCR reaction \cite{Topsoe_1994}, it is interesting to note a small body of work on the ``photo-SCR'' where anatase TiO$_2$ has been proposed as an active and selective catalyst \cite{Teramura_2002,Teramura_2003,Teramura_2004,Yamazoe_2006,Yamazoe_2007,Yamamoto_2013,Lasek_2013,Ji_2014}. These studies provide useful context for the photocatalytic nitrogen fixation as they involve the interaction of both oxidized and reduced nitrogen species with titania surfaces.

In addition to nitrogen chemistry, the interaction of titania surfaces with water will play a role in aqueous or humidified environments. Water adsorption on TiO$_2$ is a topic of extensive research and debate \cite{Sun2010,Harris2004,Kumar2013,Pang2008,Pang_2013,Benkoula2015,Lindan_1996,Bates1998,Schaub_2001,Lindan_2005}, and there are a great variety of theoretical results regarding the competition between dissociative\cite{Lindan_1996,Lindan_2005} and molecular adsorption\cite{Bates1998,Schaub_2001,Kowalski_2009} on pristine surfaces, although a clear consensus has not emerged. Analyses including the effects of slab thickness\cite{Harris2004} and method\cite{Kumar2013} support the two modes being nearly degenerate at monolayer coverages, with partial dissociation having a slightly higher energy. This viewpoint is consistent with recent experimental studies observing dissociated water on pristine surfaces.\cite{Walle2009}  
Experimental studies from the mid 1990's to the mid 2000's show that on a reduced surface (one containing vacancies at the bridging oxygens) water tends to dissociate to form two hydroxyl groups on the surface, thereby filling the vacancy \cite{Henderson1996,Schaub_2001,Krischok_2001,Ketteler_2007}, while on a pristine surface water adsorbs molecularly at the 5-fold Ti site \cite{Hugenschmidt_1994}. As previously noted there has been one recent report showing partial dissociation of water on the pristine surface.\cite{Walle2009}

Several other factors have been observed to play an important role in aqueous TiO$_2$ photochemistry. The formation of hydroxyl radicals is known to occur in both rutile and anatase \cite{Ishibashi_2000,Xiang_2011}, and these radicals participate in a number of oxidation reactions \cite{Ishibashi_2000}. Furthermore, the influence of both surface and bulk defects such as oxygen vacancies can enhance adsorption and influence photocatalytic activity \cite{Yan_2013}.
The field of TiO$_2$ photocatalysis and surface science is vast, and a thorough review is beyond the scope of this work; the reader is referred to published reports and review articles to gain a more complete view. \cite{Henderson_2011,Diebold2003,Benkoula2015,Pang2008,Pang_2013}

In this work we present an initial computational investigation of photocatalytic nitrogen reduction and oxidation on the rutile TiO$_2$(110) surface. The rutile (110) surface is hypothesized to be the active surface due to the fact that photocatalytic nitrogen fixation rates have been observed to correlate with the amount of rutile in TiO$_2$ samples \cite{Schrauzer_2011}; the (110) surface is the lowest energy surface on rutile and is likely to provide a model for other rutile surfaces. Furthermore, the rutile (110) surface has been explicitly hypothesized to be the active surface in recent experimental work \cite{Hirakawa_2017}. The availability of adsorbed nitrogen on rutile (110) is investigated under aqueous and gas-phase conditions, and the thermodynamic stability of adsorbed intermediates for nitrogen reduction and oxidation are investigated. We utilize ab initio thermodynamics \cite{Reuter_2001,Reuter_2005} and the simple but effective ``computational hydrogen electrode'' (CHE) approach \cite{Calle_Vallejo_2012} that has proven useful in understanding the oxygen reduction \cite{Norskov_2004}, oxygen evolution \cite{Man_2011}, CO$_2$ reduction \cite{Peterson_2010}, and nitrogen reduction \cite{Skulason_2012} reactions. Furthermore, the Bayesian error estimation functional (BEEF-vdW) error estimation functional \cite{Wellendorff_2012} is utilized to asses the robustness of the conclusions. The results indicate that nitrogen adsorption is more favorable under gas-phase conditions, providing a simple explanation for the improved experimental performance. However, the computational evidence refutes the hypothesis that the rutile (110) surface is the active site for nitrogen reduction due to the low stability of adsorbed N$_2$H$_{\mathrm{x}}$ and NH$_{\mathrm{x}}$ compounds at the pristine surface and at oxygen vacancy and Fe-substitution sites. In contrast, computational results indicate that nitrogen oxidation to NO is a thermodynamically viable fixation route on rutile (110), and it is hypothesized that the N-N bond is dissociated through an oxidative process and the NO intermediate is subsequently reduced.

\section{Results and Discussion}

The study of photocatalytic nitrogen fixation on TiO$_2$ is carried out using DFT calculations to provide molecular-scale insight into the thermochemistry of adsorbed intermediate states. Computational investigations of photocatalysis on TiO$_2$ are difficult due to the complex nature of photocatalytic interfaces \cite{Calle_Vallejo_2012,Hellman2017} and the electronic structure of TiO$_2$ \cite{Arroyo_de_Dompablo_2011,Diebold2003}. In this work we utilize simple models to establish the feasibility of various hypotheses. The energetics of adsorbed species are computed under ideal gas-phase conditions (e.g. solvent effects, electric field effects, and coverage effects are neglected) and it is assumed that excitation and charge transport are decoupled from the electrochemical reactions, enabling the use of the CHE approach \cite{Norskov_2004,Peterson_2010,Hellman2017,Calle_Vallejo_2012}; all voltages are relative to CHE (equivalent to RHE) unless otherwise noted. To treat the electronic structure the generalized gradient approximation (GGA) with van der Waals (vdW) level of theory is used. The use of the Bayesian error estimation functional \cite{Wellendorff_2012} (BEEF-vdW) provides error estimates that quantify uncertainty in order to assess the sensitivity of conclusions to the error due to the GGA approximation; more details are provided in the Methods section. This approach is used to test four hypotheses, as discussed in the following sections: i) photocatalytic nitrogen fixation rates are greater in the gas phase than the aqueous phase ii) rutile (110) is an active surface for nitrogen reduction iii) oxygen vacancies or Fe substititions on rutile (110) are active sites for nitrogen reduction iv) rutile (110) is an active surface for nitrogen oxidation and subsequent NO reduction.

\subsection{Nitrogen adsorption under gas and aqueous environments}

\begin{figure}
\centering
\includegraphics[width=0.8\textwidth]{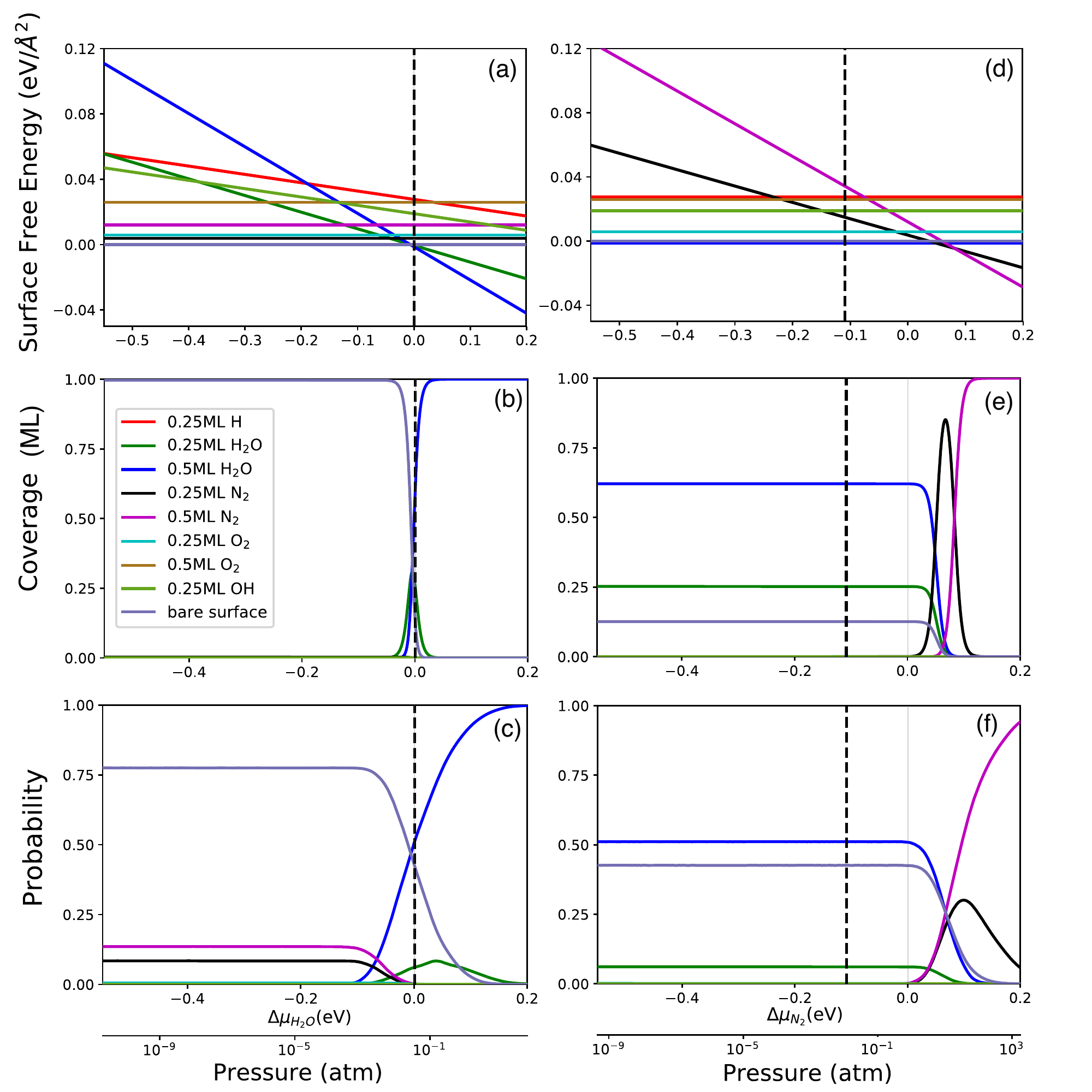}

\caption{Surface free energy (a,d), coverage (b,e), and coverage probability (c,f) for H$_2$O, N$_2$, O$_2$, and OH as a function of H$_2$O (a-c) and N$_2$ (d-f) chemical potentials. The relevant water potential under gas-phase conditions (0.035 atm) and nitrogen potential under aqueous conditions (0.012 atm) are shown by dashed lines in (a-c) and (d-f) respectively. Graphs a-c use constant N$_2$ chemical potential set at atmospheric pressure of N$_2$ (0.8 atm) and d-f use constant water chemical potential set at 100\% relative humidity of H$_2$O. Nitrogen pressure at aqueous conditions is estimated using Henry's law. The probability of various coverages (c,f) given the uncertainty in the BEEF-vdW functional is calculated using Eq. \ref{eq:prob}.}
\label{fig:surface_diagram}
\end{figure}

The literature contains examples of experiments done in aqueous solution\cite{Augugliaro_1982,Hirakawa_2017} and humidified air\cite{Schrauzer_1977,Schrauzer_1983}. Of these, humidified air generally has a higher rate.\cite{Schrauzer_2011} To better understand the difference in performance under aqueous and gas-phase conditions ab-initio thermodynamics \cite{Reuter_2001} have been used to compute surface phase diagrams for rutile (110) as a function of water and nitrogen chemical potentials. Fig. \ref{fig:surface_diagram}a shows the surface free energy (top) and surface coverage (middle) of the reactants available at ambient gas-phase conditions (H$_2$O, N$_2$, O$_2$, OH) (a similar analysis including all reactants, products and intermediates is available in the Supporting Information). The results show that the bare TiO$_2$ surface is dominant at gas-phase conditions. The ensemble of energies from the BEEF-vdW functional can be exploited to evaluate the sensitivity of these coverages to DFT error. The probability analysis (Fig \ref{fig:surface_diagram}a, bottom) suggests that there is a very low probability of having appreciable N$_2$ coverage at 100\% RH, but under dry conditions there is a probability of $\approx$ 10\% of having a N$_2$ coverage $>$ 0.25 ML. This suggests that arid environments may favor N$_2$ adsorption, though it is noted that water also acts as a proton source so nitrogen reduction cannot occur if the humidity is too low. One interpretation of the surface coverage probability is that the BEEF-vdW functional is not sufficiently accurate to precisely determine the surface coverage, though the bare surface is the most likely. Another perspective is to view this as a sensitivity analysis of coverage with respect to binding energies of competing surface species. From this view, the results indicate that adsorption sites with slightly higher relative N$_2$ adsorption are likely to have a significant N$_2$ coverage under gas-phase conditions. This could include defect sites or other facets,  although the site must have both stronger absolute binding energies and stronger relative adsorption of N$_2$ vs. H$_2$O. There are likely sites that satisfy this criterion, but their prevalence must also be considered. Regardless of the interpretation, the conclusion is that nitrogen coverages are expected to be relatively low on rutile TiO$_2$ (110) which is unsurprising given the inert nature of N$_2$.

The surface phase diagrams under aqueous conditions (\ref{fig:surface_diagram}b) indicate that water will be the dominant surface species for a wide range of N$_2$ pressures, with adsorbed N$_2$ becoming dominant at approximately 100 atm. In this case the probability analysis indicates that although the surface coverage of water/hydroxyl is not well-determined by the BEEF-vdW functional, the probability of a significant N$_2$ coverage is negligible at aqueous conditions. Interestingly, several reports have shown appreciable nitrogen fixation rates under aqueous conditions. \cite{Augugliaro_1982,Hirakawa_2017} The fact that competitive nitrogen adsorption is not favored under aqueous conditions indicates that highly reactive surface groups or defect sites with higher N$_2$ binding energy but low stability may play a role. The finding that nitrogen adsorption is more favorable in the gas phase suggests that the enhanced photocatalytic activity under gas phase conditions is due to the improved ability to adsorb N$_2$ in the absence of water. This intuitive result corroborates previous findings indicating that a key challenge in photocatalytic nitrogen fixation is getting N$_2$ to adsorb\cite{Zhu_2013,Vettraino_2002,Schrauzer_2011}, and qualitatively explains the  correlation between N$_2$ pressure and photocatalytic nitrogen reduction activity \cite{Schrauzer_2011,Ali_2016}. The issue of nitrogen adsorption must be addressed in any photo(electro)catalytic system regardless of reaction mechanism, and will be a fundamental problem for low pressure nitrogen fixation processes.

\subsection{Thermochemistry of nitrogen reduction}

\begin{figure}[h]
\centering

\includegraphics[width=0.25\textwidth]{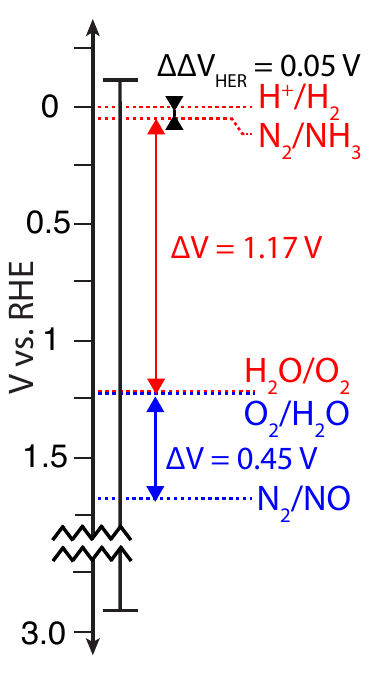}
\caption{Redox couples for nitrogen reduction (red) and oxidation (blue) with required driving force ($\Delta$V) \cite{Medford_2017}. The difference between hydrogen evolution and nitrogen reduction ($\mathrm{\Delta \Delta V_{HER}}$) is shown to highlight the selectivity challenge for nitrogen reduction. Band edges for rutile TiO$_2$ were obtained from literature \cite{Nozik_1996}  and are shown (black bar) for reference (note scale break at 1.75 V).}
\label{fig:redox_ladder}
\end{figure}

The electrochemical nitrogen reduction reaction converts nitrogen to ammonia and occurs at an equilibrium potential of 0.05 V vs. RHE (Fig. \ref{fig:redox_ladder}). The proximity of this redox couple to the hydrogen evolution reaction (H$^+$/H$_2$ at 0 V) makes selective reduction of nitrogen to ammonia challenging \cite{Skulason_2012, Montoya_2015,Singh_2017}. Furthermore, the alignment of the rutile TiO$_2$ conduction band edge (Fig. \ref{fig:redox_ladder}) indicates that the available overpotential for nitrogen reduction under photocatalytic conditions is relatively low ($<$0.15 V).  Nonetheless, selective nitrogen reduction has been observed on TiO$_2$ photocatalysts \cite{Schrauzer_1977, Schrauzer_1983,Schrauzer_2011,Augugliaro_1982}, indicating that TiO$_2$ is capable of dissociating the strong N-N bond more easily than the much weaker H-H bond. This suggests that identification of the active site for photocatalytic nitrogen reduction may enable the development of improved nitrogen reduction electrocatalysts. In this section we hypothesize that the active site is pristine rutile (110), and examine the binding free energies of intermediates for the dissociative and associative nitrogen reduction pathways.

\begin{figure}
\includegraphics[width=0.4\textwidth]{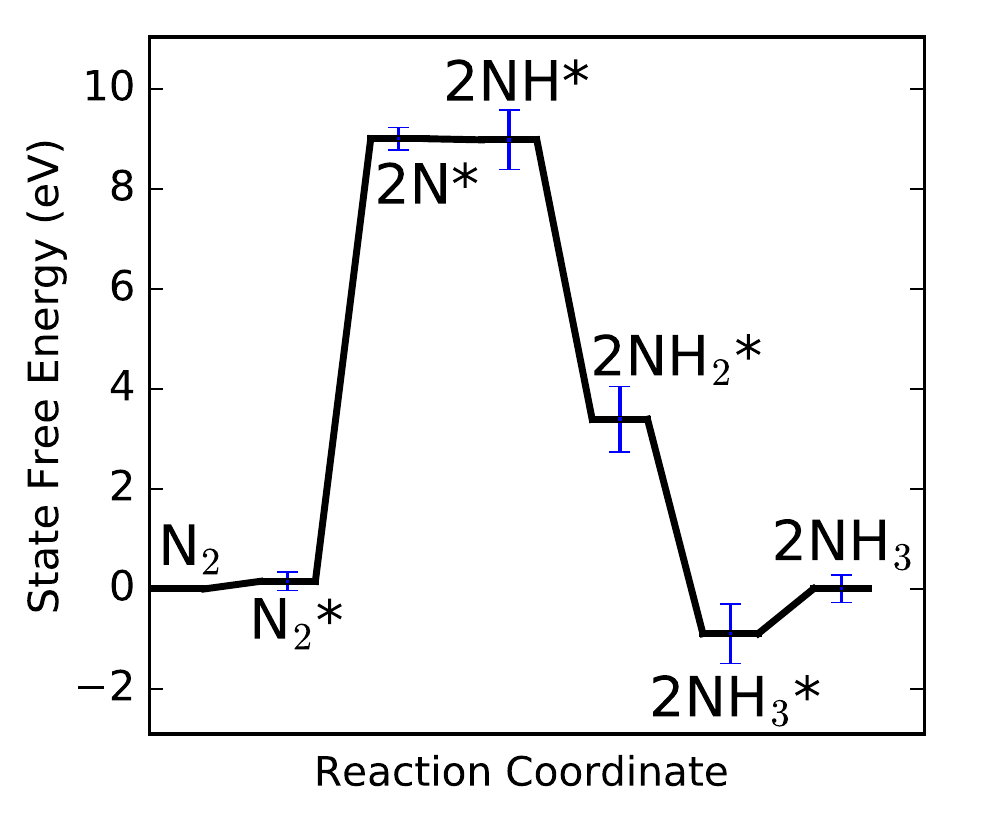}
\caption{Free energy diagram for dissociative nitrogen reduction at the equilibrium potential computed from DFT (0.008 V, compared to 0.05 V from experiment). The blue error bars represent one standard deviation of the BEEF-vdW energy ensemble. Adsorbed states are labeled, and the full reaction mechanism is listed in equations \ref{eq:diss_mech_start} - \ref{eq:diss_mech_finish}.}
\label{fig:reaction_mechanism_diss}
\end{figure}

  The mechanism for thermocatalytic nitrogen reduction (i.e. Haber-Bosch catalysis) is established as a dissociative mechanism in which the first step is scission of the N-N bond, followed by hydrogenation of adsorbed mono-nitrogen \cite{Emmett_1933,Ertl_1976,Spencer_1981,Honkala_2005}; the electrochemical equivalent is shown in equations \ref{eq:diss_mech_start} - \ref{eq:diss_mech_finish}. Calculations of N-N scission on rutile oxides shows that they follow ``ideal'' scaling, suggesting that the activation barrier for nitrogen dissociation may be low \cite{Vojvodic_2014}, hence we first investigate the dissociative mechanism. Figure \ref{fig:reaction_mechanism_diss} shows the free energy diagram for the dissociative mechanism at standard temperature and pressure at the equilibrium potential. The dissociation energy of N$_2$ is remarkably high ($>$8.5 eV), indicating that this route is not remotely thermodynamically feasible on rutile (110). This extreme barrier is not sensitive to the BEEF-vdW exchange-correlation approximation, and is expected to be prohibitively large even in the presence of solvent stabilization since each adsorbed N* would need to be stabilized by $>$3 eV, significantly more than typical solvent stabilization values\cite{Karlberg_2007,He_2017,Hellman2017}. Furthermore, this step is not electrochemically driven, so even the application of large overpotentials will not enable direct N-N scission. This provides strong evidence against direct dissociation of the N-N bond on the pristine rutile TiO$_2$ surface. The same stabilization would be required for NH* species, effectively eliminating any pathway involving NH* (e.g. dissociation of NNH). Adsorbed NH$_2$ species are somewhat more stable, and may exist under solvated conditions, opening the possibility of mechanisms involving dissociation of N$_2$H$_{\mathrm{x}>2}$ species, similar to the associative mechanism that will be discussed subsequently.
  
\begin{align}
	\label{eq:diss_mech_start}
	N_2(g)+^*  & \rightarrow N_2^{*} \\
    N_2^* + ^*  & \rightarrow 2N^*\\
    2N^* + 2(H^+ + e^-) & \rightarrow 2NH^* \\
    2NH^* + 2(H^+ + e^-) & \rightarrow 2NH_2^* \\
    2NH_2^* + 2(H^+ + e^-) & \rightarrow 2NH_3^* \\
    NH_3^*  & \rightarrow NH_3(g) + ^{*}
    \label{eq:diss_mech_finish}
\end{align}

\begin{figure}
\includegraphics[width=0.9\textwidth]{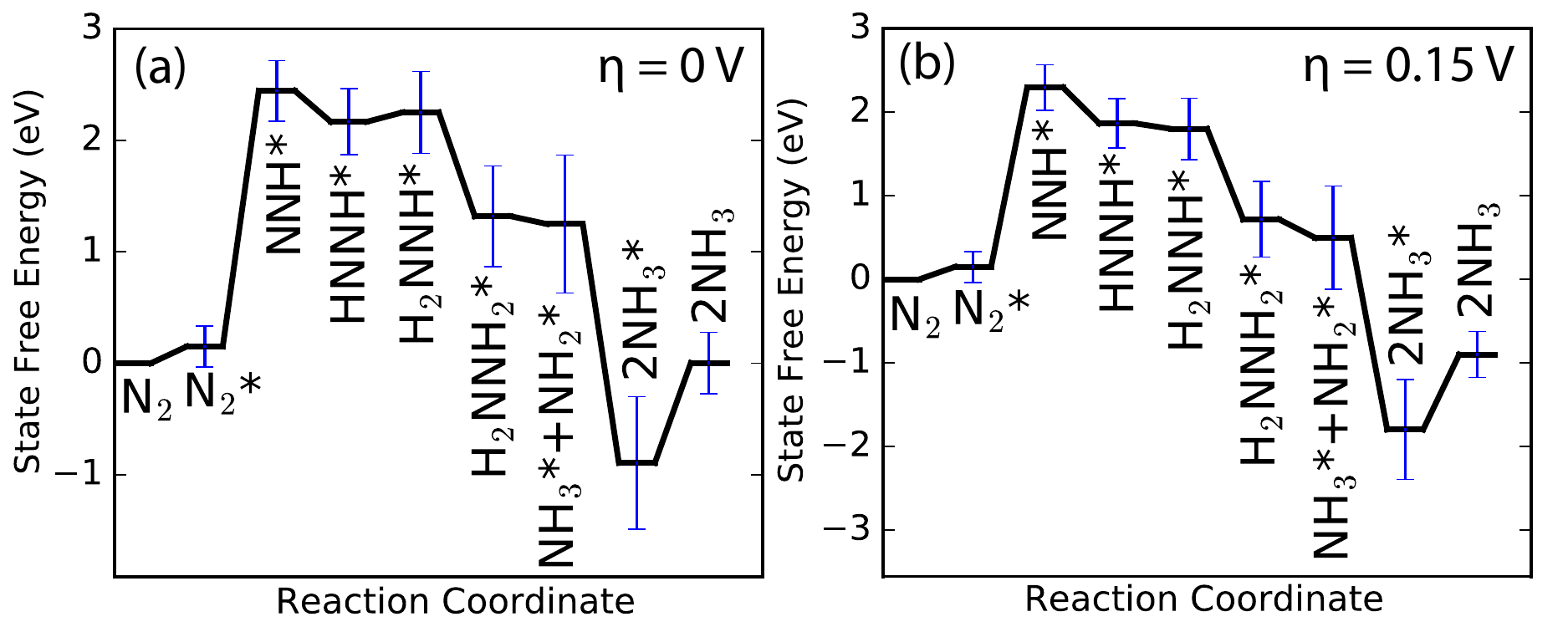}
\caption{Free energy diagram for associative nitrogen reduction at an overpotential ($\eta$) of zero (a) and at the overpotential due to the conduction band edge ($\eta=0.15$ V) (b). The equilibrium potential ($\eta=0$) is computed to be 0.008 V (0.05 V experimentally). The blue error bars represent one standard deviation of the BEEF-vdW energy ensemble. Adsorbed states are labeled, and the full reaction mechanism is listed in equations \ref{eq:assoc_mech_start} - \ref{eq:assoc_mech_finish}.}
\label{fig:FED_assoc}
\end{figure}

The associative nitrogen reduction mechanism proceeds via diazene (N$_2$H$_2$) and hydrazine (N$_2$H$_4$), as shown in equations \ref{eq:assoc_mech_start} - \ref{eq:assoc_mech_finish}. This mechanism has been proposed to be most relevant for electrochemical nitrogen reduction \cite{H_skuldsson_2017,Skulason_2012,van_der_Ham_2014}, and the fact that hydrazine has been observed as a photoreduction product on TiO$_2$ \cite{Schrauzer_2011} suggests that it may be the relevant photocatalytic nitrogen reduction mechanism. The free energy diagram for the associative mechanism at the equilibrium potential is shown in Fig. \ref{fig:FED_assoc}a, while the free energy diagram at the conduction band edge energy for rutile is shown in Fig. \ref{fig:FED_assoc}b. Under photocatalytic conditions the conduction band edge is the relevant potential, corresponding to an overpotential of 0.15 V. However, examination of the free energy diagram reveals a thermodynamic limiting potential of 2.5 V due to the unstable NNH* adsorbed intermediate, which is consistent with previous work \cite{H_skuldsson_2017}. This barrier is significantly higher than the conduction band potential, making the route improbable unless the adsorbates are stabilized significantly ($\approx$ 2 eV) by solvent/dipole effects. The rate-limiting hydrogenation of N$_2$ is consistent with studies of electrochemical nitrogen reduction on metals \cite{Skulason_2012}, nitrides \cite{Abghoui_2016}, and oxides \cite{H_skuldsson_2017} indicating a general trend for photo- and electrochemical nitrogen reduction.

\begin{align}
	\label{eq:assoc_mech_start}
	N_2(g)+^*  & \rightarrow N_2^{*} \\
    N_2^* + H^+ + e^-  & \rightarrow NNH^* \\
    NNH^* + H^+ + e^- & \rightarrow HNNH^* \\
    HNNH^* + H^+ + e^- & \rightarrow  HNNH_2^* \\
    HNNH_2^* + H^+ + e^- & \rightarrow H_2NNH_2^* \\
    H_2NNH_2^* + ^* + H^+ + e^- & \rightarrow NH_2^* + NH_3^* \\
    NH_2^* + NH_3^* + H^+ + e^- & \rightarrow 2NH_3^*  \\
    2NH_3^*  & \rightarrow 2NH_3(g) + 2^{*} 
    \label{eq:assoc_mech_finish}
\end{align}

The results indicate a prohibitively high barrier for both dissociative and associative nitrogen reduction on pristine rutile TiO$_2$ (110), although the associative pathway is significantly more favorable than the dissociative pathway. These findings are robust to the error of the exchange-correlation approximation employed, and the energetic barriers are significantly larger than the typical magnitude of electrochemical interface effects that have been neglected (solvent, electric field). This evidence refutes the hypothesis that rutile (110) is the active site for photocatalytic nitrogen reduction on TiO$_2$, necessitating the development and testing of alternative hypotheses.

\subsection{Nitrogen reduction at oxygen vacancies and iron substitutions}

Surface defects are known to play a key role in many types of heterogeneous catalysis \cite{Yates_1991}, and oxygen vacancies in particular have been shown to participate in numerous catalytic reactions on oxides \cite{McFarland_2013} including TiO$_2$ \cite{Kim2016,Kim2014,Diebold2003,Pang2008,Pang_2013}. Oxygen vacancies are typically highly reactive, leading to binding energies that are often substantially stronger than binding at the stoichiometric surface. The unstable nature of N$_2$H on pristine rutile (110) (Fig. \ref{fig:FED_assoc}) indicates that defect sites may enable nitrogen reduction by enhancing the stability of N$_2$H and other high-energy intermediates. Bridging oxygen (O-br) vacancies are known to occur commonly on rutile (110) surfaces \cite{Diebold2003,Pang2008,Pang_2013}, and a recent analysis of nitrogen reduction over titania has shown that the reaction rate is proportional to the measured number of oxygen defects.\cite{Hirakawa_2017}. Due to these considerations the O-br vacancy is a natural starting point for evaluating the effect of surface defects, although it is noted that other intrinsic defects, such as Ti vacancies in the surface or sub-surface may also play a role \cite{Hobiger_1990}.

\begin{figure}
\includegraphics[width=0.9\textwidth]{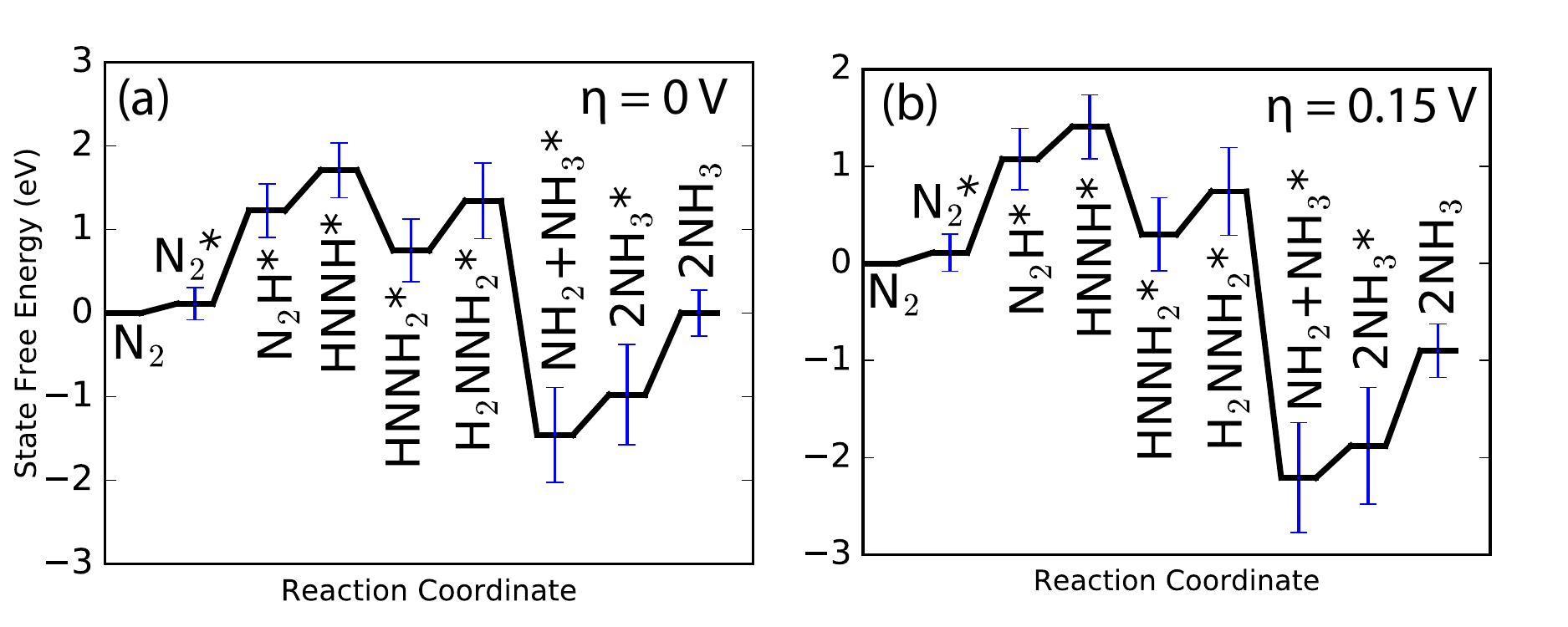}
\caption{Free energy diagram for associative nitrogen reduction at an O-br vacancy site at an overpotential ($\eta$) of zero (a) and at the overpotential due to the conduction band edge ($\eta=0.15$ V) (b). The equilibrium potential ($\eta=0$) is computed to be 0.008 V (0.05 V experimentally). The blue error bars represent one standard deviation of the BEEF-vdW energy ensemble. Adsorbed states are labeled, and the full reaction mechanism is listed in the Supporting Information.}
\label{fig:FED_defect_assoc}
\end{figure}

The energetics of the associative nitrogen reduction pathway at the rutile (110) O-br defect site are shown in Fig. \ref{fig:FED_defect_assoc}. Comparison to the energetics of the pristine surface (Fig. \ref{fig:defect_effects}a) reveals a significant stabilization of the N$_2$H intermediate, corresponding to a thermodynamic limiting potential of 1.21 V. However, examination of the energy diagram at the conduction band potential of rutile TiO$_2$ (Fig. \ref{fig:FED_defect_assoc}b) indicates that the N$_2$H and N$_2$H$_2$ intermediates are still too unstable to explain the photocatalytic reduction observed, although it is plausible that the thermodynamic barrier of 1.39 eV (energy of N$_2$H$_2$ at the conduction band edge) may be overcome by stabilization if solvent or dipole effects for N$_2$H and N$_2$H$_2$ are considerably larger than typical amounts  of $\approx$0.6 eV\cite{Karlberg_2007,He_2017,Hellman2017}. Furthermore, other defects such as sub-surface O or Ti vacancies may increase the reactivity of the surface, although as defects become less stable they will also become less prevalent on the surface. This tradeoff between stability and reactivity has been noted previously for amorphous oxides \cite{Goldsmith_2013}, and suggests that more reactive (unstable) defects will have a limited impact due to low prevalence. The O-br defect considered here has a formation free energy of 1.54 eV at 0V SHE referenced against water, indicating that it will occur with an relatively low probability on thermodynamically equilibrated surfaces. Morphological (e.g. particle edges) and kinetic (e.g. trapped bulk defects) effects will increase the prevalence of vacancies or other defects in real catalysts \cite{Yan_2013}; these effects are difficult to control and characterize, and may be the source of some discrepancies in the nitrogen photofixation literature \cite{Medford_2017}.

\begin{figure}
\includegraphics[width=0.9\textwidth]{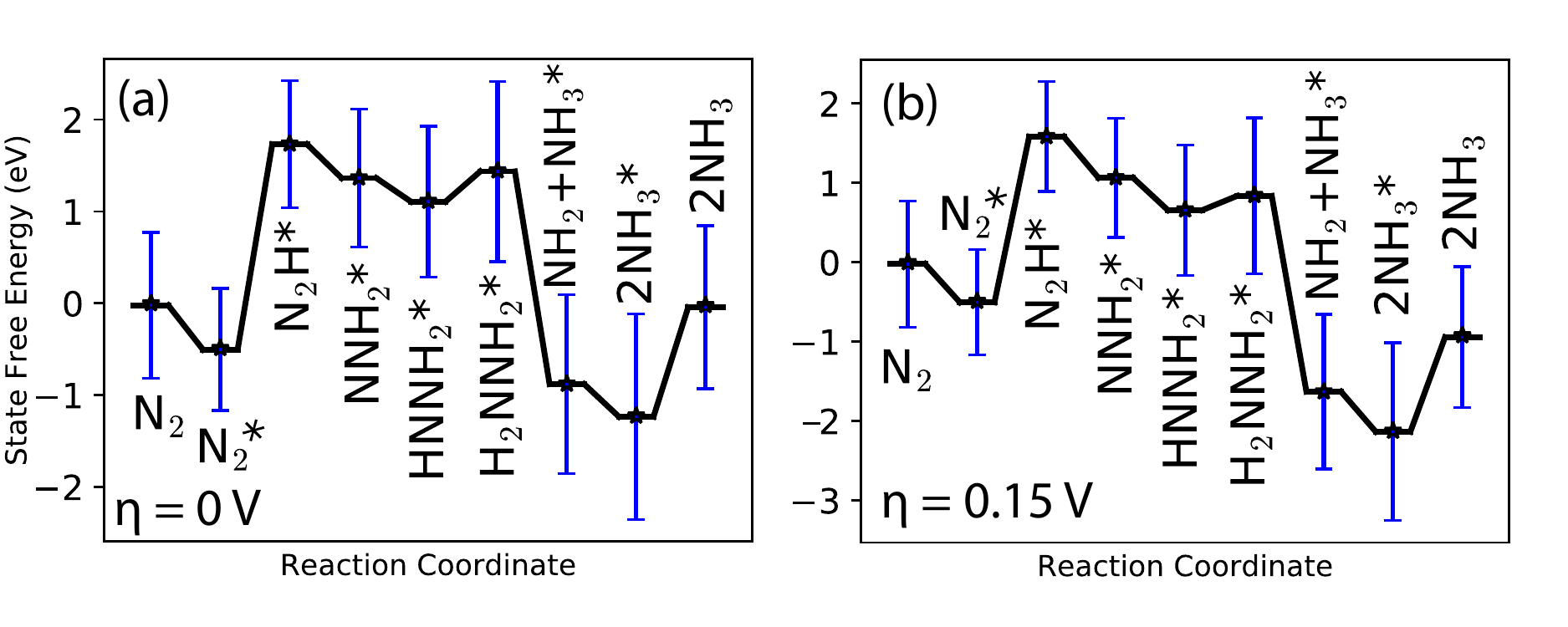}
\caption{Free energy diagram for associative nitrogen reduction at an iron substitution site at an overpotential ($\eta$) of zero (a) and at the overpotential due to the conduction band edge ($\eta=0.15$ V) (b). The equilibrium potential ($\eta=0$) is computed to be 0.008 V (0.05 V experimentally). The blue error bars represent one standard deviation of the BEEF-vdW energy ensemble. Adsorbed states are labeled, and the full reaction mechanism is listed in the Supporting Information.}
\label{fig:FED_FE_assoc}
\end{figure}

In addition, many experimental studies have noted that iron dopants significantly increase the nitrogen photofixation rates.\cite{Schrauzer_1977,Schrauzer_1983,Augugliaro_1982,Soria_1991} Iron dopants may affect catalyst activity through effects in the bulk or on the surface chemistry. In the bulk, iron has been shown to improve charge separation, reducing recombination of electron-hole pairs. This charge separation enhancement has been suggested to be the dominant effect of iron dopants for photocatalytic nitrogen fixation on titania. \cite{Soria_1991} Alternatively, an iron substitution at the surface may stabilize the states along the reductive pathway, directly improving the energetics of the process. The latter hypothesis was tested by computing the energetics of the associative mechanism on a slab with an iron substitution defect. Two defects were considered, an Fe$^{4+}$ defect arising from direct substitution of the 5-fold Ti atom, and an Fe$^{2+}$ defect formed by substitution of a Ti atom beneath a bridging O and removal of the bridging O, effectively forming a O-br defect and Fe substitution (see the Supplementary Information for visualization of the slab). The Fe$^{2+}$ defect was found to be more stable, and Fig. \ref{fig:FED_FE_assoc} shows the energetics of the associative pathway with an iron-substituted rutile (110) surface (a comparison of the free energy path for the Fe$^{4+}$ defect is available in the Supplementary Information). This mechanism is very similar to the mechanism on the pristine and O-br vacancies, with the slight difference that N-NH$_2$ is more stable than HNNH on the Fe defect. The limiting potential of 2.2 eV is comparable to that of the defected surface in Fig. \ref{fig:FED_defect_assoc} (1.7 eV), but is slightly higher due to the stronger adsorption of N$_2$. 
The energetics of the associative pathway on Fe-substitution defects are compared directly to O-br defects and pristine rutile (110) in Fig. \ref{fig:defect_effects}a, illustrating that the Fe-substitution defect has a similar effect to the O-br vacancy. The energy required to form this defect was calculated to be 1.1 eV relative to bulk rutile and BCC iron. 
This moderate formation energy is lower than that of the O-br defect, and N$_2$ adsorbs with a relatively strong binding energy of -0.5 eV. This suggests that Fe surface defects promote the formation of O-br vacancies and adsorption of N$_2$. Nonetheless, the high limiting potential of 2.2 V indicates that the Fe-substitution defect is not active for photocatalytic nitrogen reduction.

\begin{figure}
\includegraphics[width=0.9\textwidth]{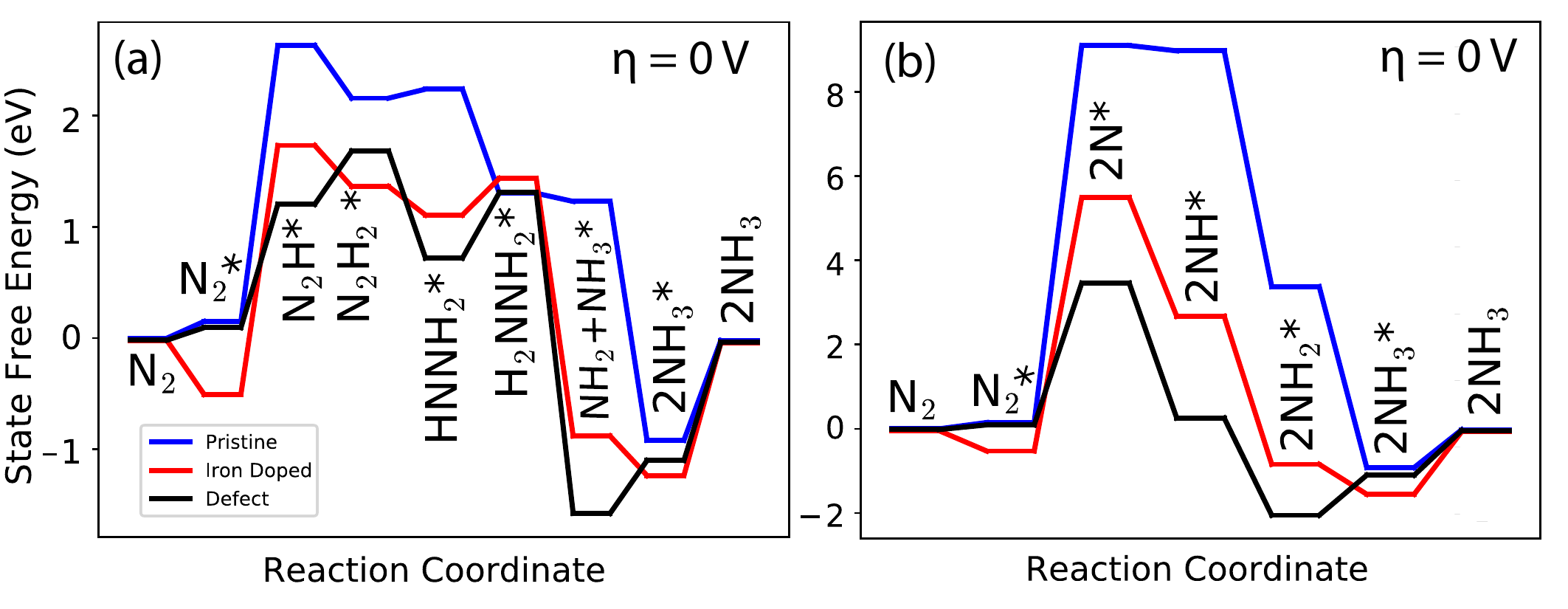}
\caption{Comparison of free energy pathways over pristine (blue), iron-doped (red) and oxygen-vacancy defect (black) for the associative (a) and dissociative (b) nitrogen reduction pathways at the equilibrium potential. All details are consistent with Figs. \ref{fig:reaction_mechanism_diss}a - \ref{fig:FED_FE_assoc}a. Error bars are omitted for clarity.}
\label{fig:defect_effects}
\end{figure}

Another possibility is that O-br vacancies and/or Fe-substitution defects significantly impact the energetics of the dissociative pathway. Iron catalysts are known to activate the N-N bond in the Haber-Bosch process\cite{Emmett_1933, Honkala_2005}, suggesting that the Fe site may play a role, and Hirakawa et. al. \cite{Hirakawa_2017} have hypothesized that direct N-N bond scission at O-br vacancies is the mechanism for photocatalytic nitrogen fixation. The energetics of the dissociative pathway for both Fe-substitution and O-br vacancy defects are compared with the energetics of the pristine surface in Fig. \ref{fig:defect_effects}b. The results show that the Fe-substitution has a relatively small effect on the thermodynamics of N-N bond scission, although a more pronounced effect is seen for NH* intermediates that are stabilized by $>$1 eV. The O-br vacancies have a much larger effect on N-N bond scission, stabilizing adsorbed N* by $>$2 eV per adsorbate. However, the thermodynamic barrier of $\approx$ 4 eV is still prohibitive at ambient conditions, and significantly higher than the 1.21 V thermodynamic limiting potential needed for the associative mechanism at the O-br vacancy. An alternative possibility is a mixed mechanism proceeding through dissociaton of partially hydrogenated species, since the NH$_{\mathrm{x}}$ species are stable at the O-br vacancy; however, this would still necessitate the formation of the potential-limiting HNNH* species from the associative mechanism and would be thermodynamically (though not kinetically) equivalent. Furthermore, we note that the free energy diagram in Fig. \ref{fig:defect_effects}b requires two O-br vacancies, since each N* (or NH$_{\mathrm{x}}$) is adsorbed at a vacancy. The relatively high formation energy of O-br vacancies suggests that this is improbable, and the experimental investigation of Hirakawa et. al. \cite{Hirakawa_2017} shows a linear dependence on oxygen vacancies, rather than the quadratic dependence that would be indicative of direct N-N (or HN-NH) scission by two vacancy sites. This finding strongly refutes the proposed hypothesis that direct N-N scission by O-br vacancies is the mechanism of photocatalytic nitrogen fixation on TiO$_2$. However, the fact that the O-br vacancy significantly stabilizes NH$_{\mathrm{x}}$ species, making NH$_{\mathrm{x}}$ binding close to exothermic suggests that it can promote nitrogen reduction and ammonia formation after the N-N bond has been cleaved.

\subsection{Nitrogen oxidation and indirect reduction}

The majority of experimental investigations of photocatalytic nitrogen fixation on TiO$_2$ catalysts have identified reduced products of ammonia or ammonium. However, several reports have observed nitrates as the main product \cite{Bickley_1979, Yuan_2013}, and examination of the standard redox potentials and TiO$_2$ band edges (\ref{fig:redox_ladder}) indicates that the band alignment for nitrogen oxidation to NO is significantly more favorable on TiO$_2$ than the band alignment for nitrogen reduction. The band alignment provides $\approx$ 1.25 V overpotential for both the oxygen reduction half-reaction and the oxidation of nitrogen for NO. Based on this we hypothesize that the scission of the N-N bond on rutile (110) proceeds via the oxidation of nitrogen to NO, which is subsequently oxidized or reduced depending on the details of the catalyst and reaction conditions.

\begin{figure}
\includegraphics[width=0.9\textwidth]{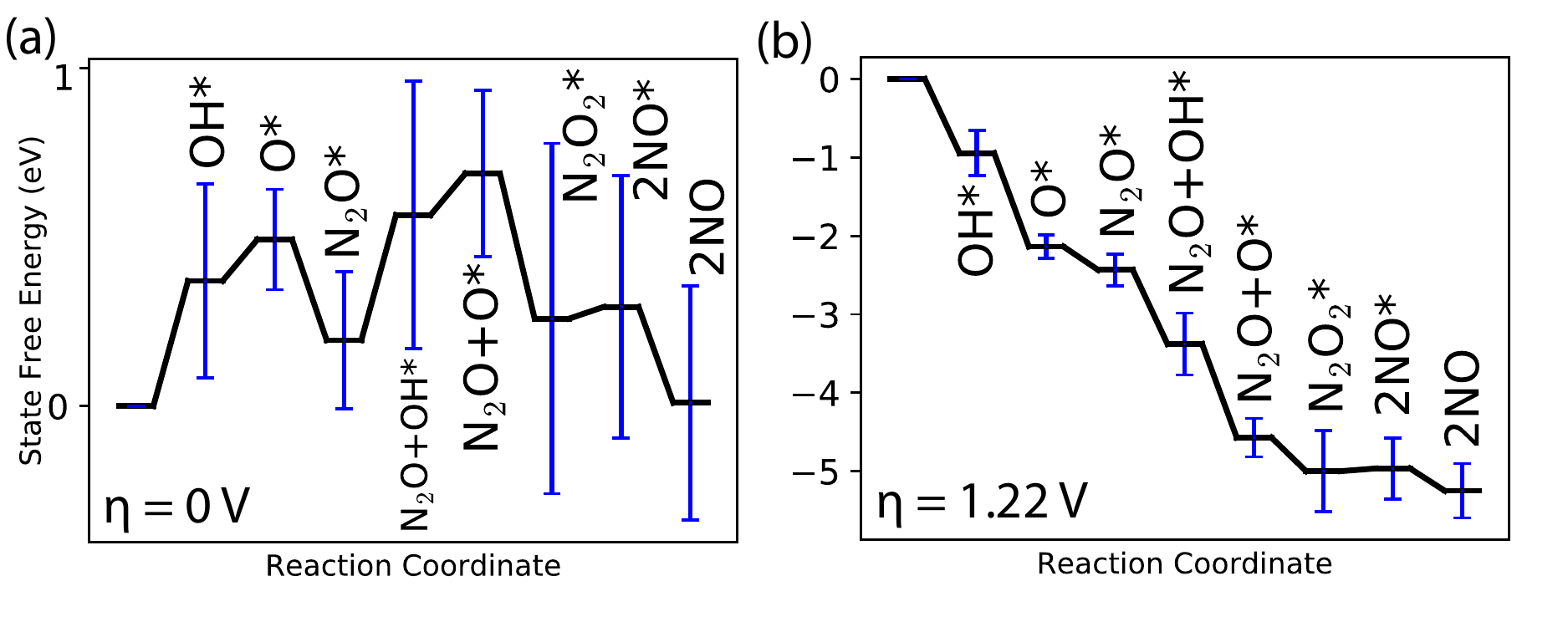}
\caption{Free energy diagram for nitrogen oxidation to NO at an overpotential ($\eta$) of zero (a) and at the overpotential due to the valence band edge ($\eta=1.22$ V) (b). The equilibrium potential ($\eta=0$) is computed to be 1.465 V (1.68 V experimentally). The blue error bars represent one standard deviation of the BEEF-vdW energy ensemble. Adsorbed states are labeled, and the full reaction mechanism is listed in the Supporting Information.}
\label{fig:FED_oxid}
\end{figure}

The thermodynamics of nitrogen oxidation intermediates on rutile (110) are shown in Fig. \ref{fig:FED_oxid} at the equilibrium potential (a) and the energy of the TiO$_2$ valence band edge (b). Examination of the free energy profile at the equilibrium potential (Fig. \ref{fig:FED_oxid}a) reveals that the thermodynamic limiting potential is 0.72 V (surface oxygen formation is the potential-limiting step), considerably lower than the case of nitrogen reduction. In addition, the direct adsorption of N$_2$ to this reactive surface oxygen is exergonic by 0.3 eV, indicating that adsorption of N$_2$ is favored by oxygen-rich surfaces. When the significant driving force provided by the photo-excited hole (1.22 V overpotential) is taken into account (Fig. \ref{fig:FED_oxid}b) the oxidative path becomes extremely favorable, with all steps being exergonic with the exception of N$_2$O$_2$ dissociation which is very slightly ($<0.1$eV) uphill. We note that this analysis considers only the thermodynamics of adsorbed states, and neglects activation barriers that will ultimately govern the kinetics of the process. These barriers could potentially be significant for N-O coupling, although the strong driving force under photocatalytic conditions will improve the kinetics of any electrochemical step.

\begin{figure}
\includegraphics[width=0.9\textwidth]{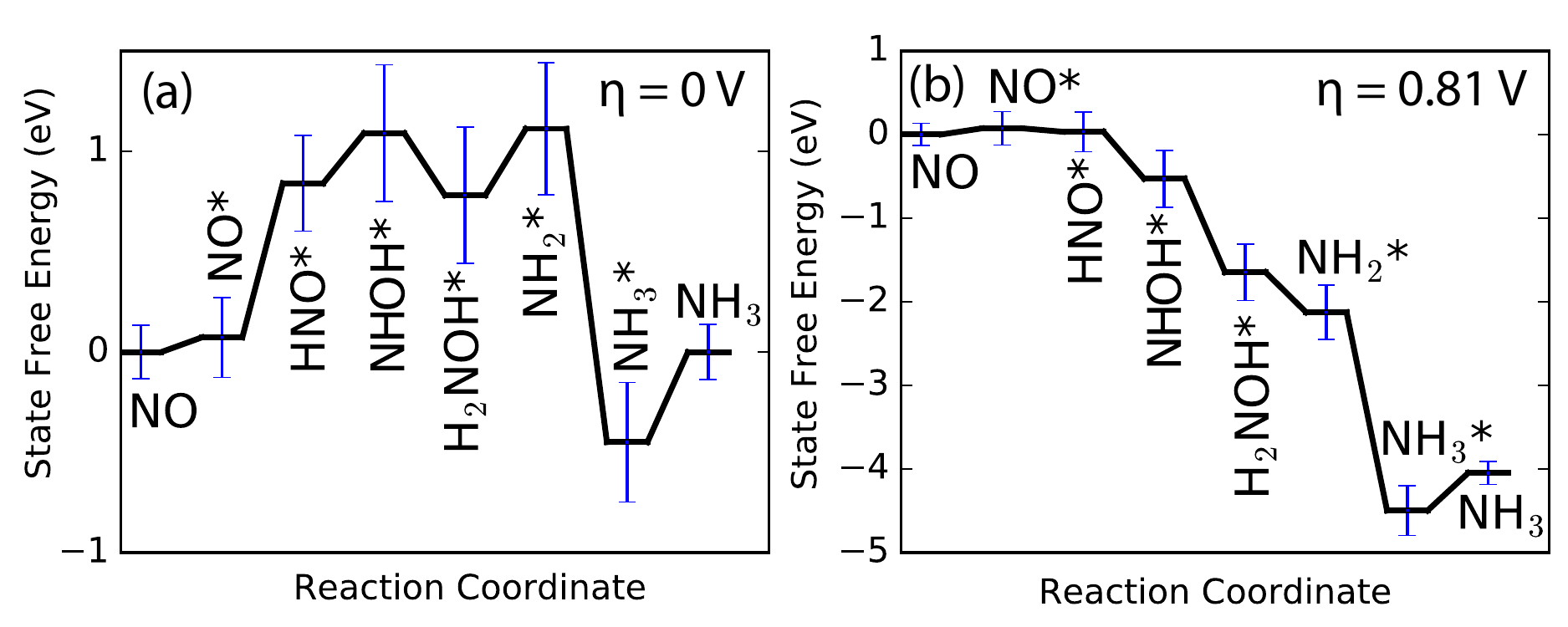}
\caption{Free energy diagram for NO reduction to NH$_3$ at an overpotential ($\eta$) of zero (a) and at the overpotential due to the valence band edge ($\eta=0.81$ V) (b). The equilibrium potential ($\eta=0$) is computed to be 0.59 V (0.71 V experimentally). The blue error bars represent one standard deviation of the BEEF-vdW energy ensemble. Adsorbed states are labeled, and the full reaction mechanism is listed in the Supporting Information.}
\label{fig:FED_oxid_red}
\end{figure}

The computational results provide strong evidence that nitrogen oxidation is thermodynamically feasible on the rutile (110) active site, while N$_2$ reduction is thermodynamically challenging. This is at odds with the experimental observation of reduced products on TiO$_2$ \cite{Schrauzer_1977,Schrauzer_1983,Augugliaro_1982,Soria_1991,Schrauzer_2011,Hirakawa_2017}. One possible explanation is that nitrogen is first oxidized to NO and subsequently reduced to ammonia. The conversion of NO to NH$_3$ is a 5 electron process with a redox potential at 0.71 V vs. RHE, well below the band gap of TiO$_2$\cite{Medford_2017}, and has been reported experimentally\cite{Ranjit_1997} and studied theoretically\cite{Xie_2017}. The thermodynamic feasibility of this pathway on rutile (110) has been computed and the most thermodynamically favorable path is shown in Fig. \ref{fig:FED_oxid_red}. The mechanism and energetics are consistent with prior work\cite{Xie_2017}, and indicate that this is indeed thermodynamically feasible. However, reduction of nitrogen oxides is a complex process that can also form partially reduced species such as N$_2$O or N$_2$. In particular, the reaction of NO to N$_2$O and the reaction of N$_2$O to N$_2$ has been observed under UHV conditions by Yates and colleagues\cite{Rusu2000,Rusu2001}. Fully understanding the selectivity of photocatalytic NO reduction on TiO$_2$ is beyond the scope of this work, but selectivity should be considered in future studies of NO reduction.


Experimentally, titania photocatalysts have been reported to reduce nitrates under aqueous conditions, although selectivity to dinitrogen vs. ammonia varies widely based on preparation conditions and metal dopants \cite{Kobwittaya_2014, Kominami_2010, Ranjit_1997,Shand_2013,Xie_2017,Lozovskii_2009}. Contrarily, oxidation of ammonia has also been reported for TiO$_2$ photocatalysts \cite{Pollema_1992, Wang_1994, Kominami_2014}, as well as simultaneous reduction of nitrate and oxidation of ammonium to N$_2$ \cite{Kominami_2014}. Furthermore, the observation of hydrazine as a product from TiO$_2$ nitrogen photofixation is not explained by this mechanism \cite{Schrauzer_2011}. These conflicting results suggest that other hypotheses should also be considered. In particular, the formation of hydrazine should be examined more closely to determine if hydrazine could be formed through recombination of NH$_x$ species (consistent with the oxidative pathway) or if it is formed associatively (consistent with the reductive pathway). Isotopic scrambling experiments\cite{Urabe_1978} and use of photoelectrochemical applied bias experiments may provide insight into this outstanding question.

\section{Conclusions}

The hypothesis that any of the sites on pristine rutile (110) are the active site for nitrogen reduction was shown to be false based on the results of BEEF-vdW DFT calculations for both the associative and dissociative mechanism, although the associative mechanism was found to be considerably more thermodynamically favorable. The revised hypothesis that oxygen vacancies or Fe substitutions are active sites was shown to be more plausible, and a significant stabilization of NH$_{\mathrm{x}}$ by O-br vacancies was found. Yet, the predicted thermodynamic barriers are considerably higher  than the conduction band edge, and the defects are not predicted to be thermodynamically stable under operating conditions, leading to the conclusion that this hypothesis is also improbable. However, the results indicate that N-N bond cleavage is thermodynamically facile on rutile (110) through an oxidative pathway, particularly with the strong oxidative driving force provided by photogenerated holes. Based on this finding, the hypothesis of an oxidized NO* intermediate in nitrogen reduction is introduced as a possible mechanism for photocatalytic nitrogen fixation on rutile (110). The work provides initial molecular-scale insight into the mechanisms that underly photocatalytic nitrogen fixation on TiO$_2$ by conclusively eliminating several possible explanations for this important process and identifying a novel hypothesis of indirect reduction through an oxidized intermediate.

\section{Methods}
\label{sec:methods}
\subsection{Computational}
\subsubsection{Density Functional Theory}
In the current work the Quantum ESPRESSO software package\cite{QE-2009} is used in conjunction with the Atomic Simulation Package (ASE)\cite{ISI:000175131400009} to carry out plane wave density functional theory calculations. The BEEF-vdw functional \cite{Wellendorff_2012} is used for all work with a plane wave cuttoff of 400 eV and a 4x4x1 Monkhorst-Pack k-point grid is used for all slab models \cite{Monkhorst_1976}. Spin polarization and a dipole correction\cite{Bengtsson_1999} (applied along the axis of the slab) are used for all slab calculations. Structures are optimized using the BFGS line search method to a total force of 0.05 eV/\AA{}. The geometries and adsorption energies for all species are available in Tables S1 and S3. All gas-phase species are evaluated in a periodic unit cell with 6 \AA{} of vacuum space and a $\Gamma$-point k-point sampling. A number of calculated properties of TiO$_2$ (e.g. band gap, water adsorption energy) vary with the number of layers in the model slab \cite{Sun2010,Bredow2004,Patel2014}; slabs with an even number of layers are known to be more accurate due to a symmetric electronic environment between sets of layers \cite{Sun2010}. The error due to slab size is expected to be ca. 0.2 eV \cite{Sun2010}, similar to the error of DFT for adsorption energies \cite{Wellendorff_2015}. Furthermore, there is debate regarding the accuracy of GGA functionals for TiO$_2$. Several studies have indicated that catalytic properties, even of defects, can be reliably treated with GGA functionals \cite{Schaub_2001,Li_2008,Zheng_2016}, although other work indicates that hybrid, +U, or double-hybrid methods are necessary, particularly to treat electronic transitions and optical properties  \cite{Morgan_2007,Landmann_2012,De_k_2011, Jauho_2015}. In this work we utilize the error estimation capabilities of the BEEF-vdW functional \cite{Wellendorff_2012} in order to estimate its accuracy and quantify the uncertainty due to the GGA approximation. Although the quantitative accuracy of BEEF-vdW error estimation in oxide materials is not well-known \cite{Walker_2016}, the BEEF-vdW ensembles provide a systematic way to assess the sensitivity of conclusions to accuracy of approximations \cite{Medford_2015a}. The error bars shown in this work correspond to $\pm 1 \sigma$ of the ensemble of 2000 energies provided by the BEEF-vdW functional. The uncertainty quantified in the BEEF ensembles is also propagated to phase diagrams (Fig. \ref{fig:surface_diagram}). The probability of a species on the surface is defined as:
\begin{align}
	P_i = \frac{1}{N}\sum_{l} ^{N} \frac{\exp(\frac{-G_i^l}{kT})}{\sum_{j} ^{M}\exp(\frac{-G^l_j}{kT})}
	\label{eq:prob}
\end{align}
where $P_i$ is the probability of species $i$ given DFT uncertainty, $G_i^l$ is the free energy of species $i$ computed from energy $l$ of the BEEF-vdW ensemble, $M$ are the number of total species considered, and $N$ is the total number of energies in the BEEF-vdW ensemble (2000). This is equivalent to the average surface coverage from a ensemble of phase diagrams generated from the energies of the BEEF-vdW ensemble. 

\subsubsection{Thermochemistry}

Ground state electronic energies ($E_{ele}$) are obtained from DFT and converted to free energies by including zero point energy (ZPE) and thermal contributions:
\begin{align}
	G_i^o &= E_{ele} + E_{ZPE} +  \Delta H - T \Delta S \label{eq:G_ref}
\end{align}
where $E_{ZPE}$ is the zero-point energy and $\Delta H$ (enthalpy) and $\Delta S$ (entropy) are thermal contributions computed using the vibrational frequencies computed through a finite difference approximation to the Hessian. Adsorbates are treated using the harmonic approximation with a low-frequency cutoff of 30 cm$^{-1}$, while gas phase molecules were treated as ideal gases. The ASE implementation \cite{ISI:000175131400009} is used for vibrational analysis and statistical mechanics corrections. Vibrational frequencies for all species are reported in Table S2. Vibrational frequencies of adsorbates at defect sites are assumed to be the same as vibrational frequencies on the pristine surface. All thermodynamics were evaluated at 300 K and gas partial pressures were set approximate atmospheric conditions (0.8 atm N$_2$, 0.2 atm O$_2$). Liquid water was approximated as an ideal gas at saturation pressure at 300 K (0.035 atm). Relative (formation) energies were computed with respect to reference states using the equation:
\begin{align}
	G_i = G_i^o - \sum_j n_j \mu_j\label{eq:G_sum}
\end{align}
where $G_i$ is the formation free energy of species $i$, $G_i^o$ is the ``absolute'' free energy computed from DFT and free energy corrections (see Equation \ref{eq:G_ref}), $n_i$ is the number of atoms $j$ in species $i$, and $\mu_i$ is the reference chemical potential. The reference for nitrogen is N$_2$ ($\mu_N = \frac{1}{2} G_{N_2}^o$), the reference for oxygen is H$_2$O ($\mu_{H_2O} = \frac{1}{2} G_{H_2O}^o - \mu_H$) and the reference for hydrogen is set by the computational hydrogen electrode ($\mu_H = \frac{1}{2} G_{H_2}^o$ + eU). In this work no gas-phase corrections (e.g. O$_2$ \cite{Norskov_2004}, CO$_2$ \cite{Peterson_2010}) are applied since it is difficult to rigorously correct the BEEF-vdW ensembles; the quantification and propagation of error with the BEEF-vdW ensembles will capture these errors to some extent, and overpotentials are defined relative to the equilibrium potential computed by DFT in order to minimize the influence of inaccurate equilibrium potentials from DFT (see following section).

\subsubsection{Photoelectrochemistry}
\label{sec:PEC}

The photoelectrochemical model used assumes that excitation, charge transport, and charge transfer are decoupled from electrochemistry and is largely consistent with the approach reviewed by Hellman and Wang \cite{Hellman2017}. All states involving electrons utilized the CHE model reaction as a reference electrode. This model sets the free energy of reaction of hydrogen splitting, reaction \ref{eq:HER}, to zero at a potential of zero. This ``computational hydrogen electrode'' (CHE) is conceptually equivalent to the reversible hydrogen electrode (RHE)\cite{Peterson_2010}.

The energy of each state involving the addition of an electron is varied by an energy of $eU$, where e is the fundamental change and U is the potential relative to this zero point, at potentials other than zero reaction:

\begin{equation}
	\label{eq:HER}
	\frac{1}{2}H_2\rightarrow H^{+}+e^{-}
\end{equation}

This model was also used to assess the effect of photo-excited electrons and holes. The locations of the band edges were obtained from the literature \cite{Nozik_1996} and used to set the potentials of the electrons and holes relative to a the redox couples of the relevant reactions so that the overpotential is equivalent to the experimentally expected overpotential. This approach causes the \textit{absolute} potential of electrons/holes to depend on the reaction in question (due to DFT errors in the reaction energies), while the \textit{relative} energy of electrons/holes to a given redox couple is equal to the experimental value. The computed (and experimental) equilibrium potential of a given reaction is provided in the caption of each figure.

\subsubsection{Determination of adsorption sites}

Several adsorption sites were tested for stability for all stable molecular species and adatoms. For most stable molecular species the only viable adsorption site was the 5-fold titanium site; attempts to adsorb species to the bridging oxygen or in plane oxygens either led to desorption or reorientation to the 5-fold titanium site. These findings are in agreement with previous first-principles examinations.\cite{Sorescu2000,Stodt2013} Adatoms (N*, O*, H*) were found to have multiple adsorption sites. Oxygen was found to adsorb to the 5-fold titanium site or form a diatomic species at the bridging oxygen site. The latter can be idenified as the $\alpha$-oxygen species reported by Lu.\cite{Lu1994} Hydrogen adatoms were found to be most stable at the in plane oxygen, and the nitrogen adatom was found to have highest stability over the bridging oxygen. A full list of energies and structures is provided in the Supplementary Information.

\begin{acknowledgement}

The authors thank the School of Chemical \& Biomolecular Engineering 
at the Georgia Institute of Technology for providing start-up funding
for this work. The authors are also grateful to Marta C. Hatzell for constructive discussions, and Timothy Doane for providing references regarding photocatalytic nitrogen fixation.

\end{acknowledgement}
\section{Publication Disclosure}
This document is the unedited Author's version of a Submitted Work that has been submitted for publication in ACS Sustainable Chemistry and Engineering, copyright © American Chemical Society.


\begin{suppinfo}
The Supporting Information is available and contains:

All adsorption energies, vibrational frequencies, and structure images; surface free energy, coverage, and probability diagrams including all species under oxidizing and reducing conditions; reaction equations for all mechanisms and defect formations; comparison of free energy pathway for different Fe substitution models.

\end{suppinfo}

\bibliography{main} 
\newpage
\section{For Table of Contents Use Only.}
\begin{figure}

\includegraphics[width=8.47cm,height=4.76cm,]{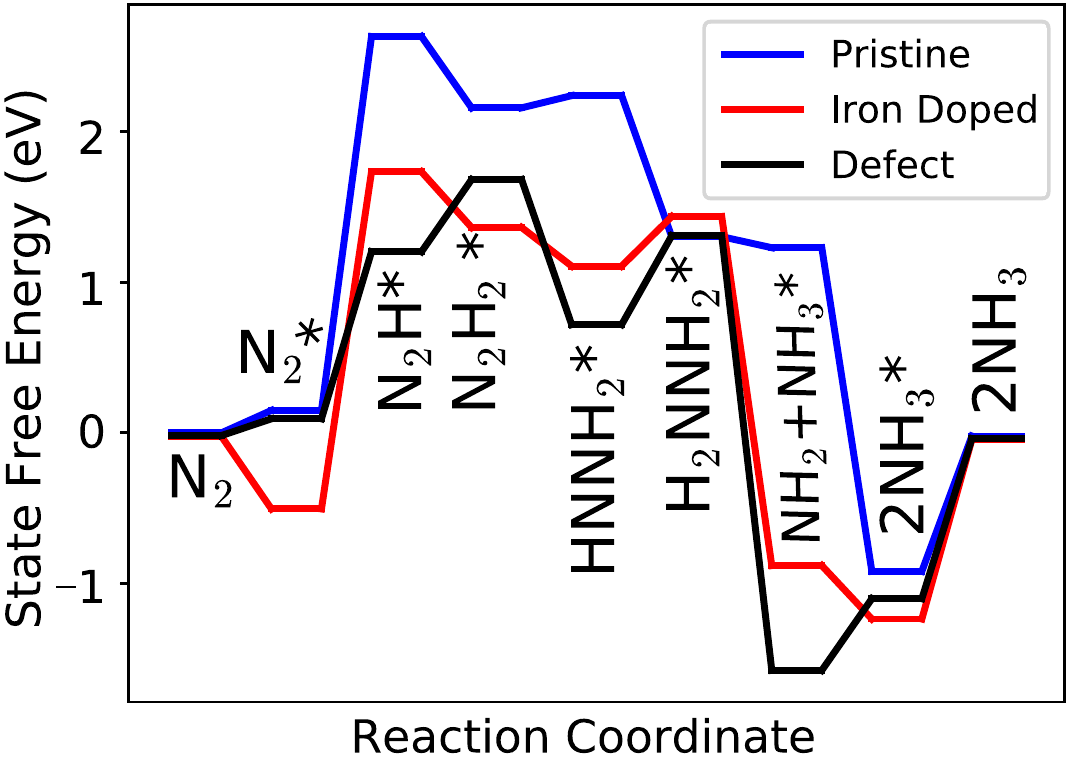}
\caption{Energetics of photocatalytic nitrogen fixation over rutile (110) titania active sites. Photocatalysis provides a route to nitrogen fixation at atmospheric conditions and sustainable fertilizer production.}
\label{fig:TOC}
\end{figure}

\end{document}